\documentclass[aps,preprint]{revtex4}
\usepackage{epsfig}
\usepackage{amsmath}
\usepackage{epsfig}
\begin{document}
\title
{\bf Spectral singularity  and deep multiple minima in the reflectivity in non-Hermitian (complex)  Ginocchio potential}    
\author{Ananya Ghatak$^{*,a}$, Bhabani Prasad Mandal$^{*,b}$ and Zafar Ahmed$^{{**},c}$ }  
\email{a:gananya04@gmail.com, b:bhabani.mandal@gmail.com, c:zahmed@barc.gov.in}
\affiliation{$~^*$Department of Physics, Banaras Hindu University, Varanasi, 221005, India\\
$~^{**}$Nuclear Physics Division, Bhabha Atomic Research Centre,
Mumbai 400 085, India}

\date{\today}

\begin{abstract} We bring out the existence of at most one spectral singularity (SS) and deep multiple minima 
in the reflectivity of the non-Hermitian (complex) Ginocchio potential. We find a parameter dependent single spectral singularity in 
this potential provided the imaginary part is emissive (not absorptive). The reflectionlessness of the real Hermitian
Ginocchio's potential at discrete positive energies gives way to deep multiple minima in reflectivity when this potential is 
perturbed and made non-Hermitian (complex). A novel co-existence of a SS with deep minima in reflectivity is also revealed
wherein the first reflectivity zero of the Hermitian case changes to become a SS for the non-Hermitian case. 
\end{abstract}

\maketitle

Scattering from a complex potential accounts for the disappearance of incident flux of particles into unknown channels.
The complex potentials are often known as optical models in nuclear and condensed matter physics. Generally, the imaginary part
of these optical models is taken to be a negative definite function of the space dimension to study the scattering with absorption. 
With the discovery [1] that the non-Hermitian complex PT-symmetric potentials (invariant under the joint transformation of Parity and 
Time-reversal) may too have real discrete spectrum, the interest in the scattering from complex potentials has been re-newed [2,6].

The reflectivity of a complex potential has been proved to depend upon the side (left/right) of the incidence of the particle if
the total complex potential is spatially asymmetric. This phenomenon is known as handedness [2,3] of the reflectivity whereas the transmittivity 
has been proved to be invariant of left/right incidence. The complex PT-symmetric potentials essentially have the imaginary part as 
anti-symmetric, therefore the handedness of the reflectivity turns out to be the essential feature of scattering from such potentials [3].
Subsequently as the complex PT-symmetry is described in terms of pseudo-Hermiticity, the handedness of the reflectivity also becomes the featureof pseudo-Hermitian interactions [4].

The Jost function $f(k)$ in case of self-adjoint Hamiltonians is known  not to vanish at a real $k$. However, for the non-self-adjoint Hamiltonians  
the Jost function may vanish at real values of $k=k_{*}$, this energy ($E_{*}=k_{*}^2$) is called spectral singularity (SS) [7]. Theoretically it is debated as to whether spectral singularities cause real discrete spectrum of a complex PT-symmetric to be incomplete [8].
 Later it has been
proved that at these real positive energies both the Reflection ($R(E)$) and Transmission ($T(E)$) co-efficients become infinite [9]. These large peaks
in the reflectivity and the transmittivity have been found akin to the zero-width resonances observed in the wave propagation in the waveguides [10]. 
Owing to the connection of a SS with infinite reflection and transmission, there is a remarkable progress in the experimental realization of the spectral singularities [11]. 

The knowledge of spectral singularities in a complex potential (PT- symmetric or non-PT-symmetric) becomes worth while to
understand spectral singularity more deeply.  
The  Scarf II potential turns out to be the first exactly solvable model of complex PT-symmetric potential having a spectral singularity.
The parametric condition on potential parameters and the exact explicit energy $E_{*}$  have been derived [12] by finding the positive energy 
pole in $T(E)$ and $R(E)$. Notably in this model the single SS results irrespective of the fact whether its real part is a barrier or a well. 
Later, when the real part of Scarf II is a well the spectral singularity have been re-obtained [13].
Next, a very interesting new parametrization of Scarf II has been proposed. According to this work one and two spectral singularities can exist
in the complex PT-symmetric and non-Hermitian cases, respectively. Moreover, several (in)variance of the reflection and transmission probabilities 
$(R(k), T(k))$ for the scattering from non-Hermitian potentials have been proposed [14]. The most recent ideas such as
unidirectional invisibility [15,16] self dual spectral singularities [17] can be seen to conform to the (in)variances proposed in [14]. Recently the
phase-space trajectories of complex PT-symmetric Scarf II potential at the spectral singularity [14] have been found to have new features [18]. 

More recently the spectral singularities of a complex non-PT-symmetric potentials are being discussed to be even  more useful in laser optics for the conceptualization of new kinds of lasers [17].
In this regard it becomes important to study various solvable models and conditions under which complex potentials give rise to spectral singularity. Specially the  count  (one, two or more) [12] of the spectral singularities requires case studies. In this Letter we would like to present the non-Hermitian complex version of the versatile 
Ginocchio potential [19] having at most one spectral singularity and several other interesting features therein. 
Presently, we consider the complexified yet non-PT-symmetric extension of the Ginocchio's potential.

The real Hermitian Ginocchio's potential [19,20] is written as 
\begin{eqnarray}
&V(x)=\mp \lambda^2 \nu(\nu+1) (1-y^2) +{1-\lambda^2 \over 4} [5(1-\lambda^2)y^4-(7-\lambda^2)y^2+2](1-y^2),
\end{eqnarray}
where
\begin{eqnarray}
&x=\lambda^{-2}[\tanh^{-1} y-(1-\lambda^2)^{1/2} \tanh^{-1}(1-\lambda^2)^{1/2} y], x\in (-\infty; \infty), y\in (-1;1). 
\end{eqnarray}
By defining 
\begin{equation}
k(E)={\sqrt{2mE} \over \hbar}, \mu(E) =\sqrt{1/4 \pm \nu(\nu+1)+ {\lambda^2-1 \over \lambda^4}k^2}-1/2.
\end{equation}
By solving the time-independent one-dimensional Schr{\"o}dinger equation, the exact transmission and reflection amplitudes are derived [19,20] as
\begin{equation}
r(k)={\Gamma[ik/\lambda^2] \Gamma[\mu +1 -{ik/\lambda^2}] \Gamma[-\mu-ik /\lambda^2] e^{2ikr_1} \over \Gamma[-ik/\lambda^2] \Gamma[\mu+1]\Gamma[-\mu]}
\end{equation}
\begin{equation}
t(k)= {\Gamma[\mu+1 -{ik/\lambda^2}]\Gamma[-\mu - ik /\lambda^2] e^{2ikr_1} \over \Gamma[-ik/\lambda^2] \Gamma[1-ik/\lambda^2]}
\end{equation}
The reflectivity and the transmitivity are given as $R(E)=|r(k)|^2$ and $T(E)=|t(k)|^2$, respectively.
This exactly solvable potential is very versatile as it can represent a thin or wide attractive potential well, a well with two side barriers and 
a repulsive potential barrier. It yields analytic and explicit expressions for discrete real (bound states)  and complex (resonant) eigenvalues.
This potential is also an interesting model of reflectionlessness (transmission resonances) [19,20] at real discrete energies. It may be remarked that reflectionlessness of a Hermitian potential is a well studied phenomenon in the literature [19-21].
 
We propose to complexify the parameter $\nu$ and leave $\lambda$ as real in (1) in order to make it complex non-Hermitian.
We then bring out the existence of at most one spectral singularity in this non-Hermitian potential in various cases. We show that the reflectionlessness of the Hermitian model (1) at discrete energies gives way to deep minima 
in reflectivity [19,20] even when this model is perturbed to become non-Hermitian.

The spectral singularities are the positive energy poles of $r(E)$ and $t(E)$ [9], we have to extract such poles from Eqs.(4,5).
Similarly we shall be extracting the reflectivity-minima from (4).
Unlike the earliest exactly solvable Scarf II the spectral singularity and the parametric condition for its existence can not be derived 
here exactly, analytically and explicitly from Eqs (4,5). In this regard we have to carry out a numerical investigation.

For a fixed non-real values of $\nu$ and real value of $\lambda$ the common condition for real energy pole of $r(E)$ and $t(E)$ is given by
\begin{equation}
\Delta(E)=-\mu(E) - ik(E) /\lambda^2= F(E)+iG(E) =n, n \in I^-+\{0\}.
\end{equation}
At this condition the Gamma function becomes infinity namely $\Gamma[n]=\infty$, here  $F(E)$ and $G(E)$ are real. 
First we fix $\nu$ and vary $\lambda$ (or vice versa). For appropriate choices of parameters, we see that the curve of $G(E)$ cuts the energy axis at some point.
We may then see that curve $F(E)$ admits a value close to a negative integer around this energy point. This is how we locate the spectral singularity, $E_*$, approximately. We then improvise our calculations around these parametric values to locate the spectral singularity exactly. This process ensures at most one spectral singularity.

Let us define 
\begin{equation}
\Omega(E)=\mu(E)+1-ik(E)/\lambda^2= H(E)+iJ(E).
\end{equation}
To explore the possibility for the second spectral singularity we plot $H(E)$ to find that it may or may not cut the energy axis, nevertheless $H(E)$ remains positive definite thereby ruling out the possibility of a second spectral singularity. So unlike non-Hermitian Scarf II [12,14] this potential can not entail a second spectral singularity.

We take $2m=1=\hbar^2$, in our calculations and summarize the results in Table 1.
A spectral singularity in Ginocchio's complex non-Hermitian potential [Eq.(1)] occurs when the real part is a repulsive barrier (see ($\{1-9\}$) 
or an attractive well (see $\{10-13\}$) or a well with side barriers (see $\{14-18\}$). Also a SS can occur irrespective of the fact whether the strength of imaginary 
part is more (see $\{3,4,6,8-11,14-17\}$) or less (see $\{1,2,5,7,12,13,18\}$) than that of the absolute value of the real part. 
Notice that in the cases when the  real part is a barrier (see $\{1-9\}$), a SS occurs at energy above the barrier height. 
The cases ($\{17,18\}$) entail the co-existence of  one SS and several minima in the reflectivity (see Fig. 3(a)).
In all the cases, a SS occurs when the imaginary part is positive (emissive) and does not seem to occur when it is negative (absorptive). Here the cases (see $\{19,20\}$) underline this observation.
	  
The case $\{11\}$ of spectral singularity is displayed in Fig. (1). A large peak in both $R(E)$ (light, dotted curve) and $T(E)$ (dark curve) at 
$E=E_*=166.720$. The shapes of the real and imaginary parts of $V(x)$ (1) are given in Fig. 1(b). Here the real part is a well. Notice that in Fig. 1(c)
$G(E)$ is monotonically decreasing and cuts the energy axis at $E=E_*$ at this energy $F(E)$  cuts the line $y=-9$, fulfilling
the condition (6), namely $\Delta(E)=-9 $ and $\Gamma(-9)=\infty$ and hence a spectral singularity. But on the contrary in Fig. 1(d),
$J(E)$ is decreasing and negative, whereas $H(E)$ is increasing and positive ruling out the existence of a second SS. 

In Fig. (2) we display a typical scenario of the reflectionlessness of the Hermitian case ($\nu=-1/2 \pm 2i$, light curves) 
and the energy oscillations in the reflectivity for the non-Hermitian case (dark curves) when $Re(\nu) \approx -0.5$
and $Im(\nu)$ is small. Smaller the $Im(\nu)$  sharper are the reflectivity-minima, but here we take a moderate value of 2.
For all the cases (a-d) in this figure $\lambda=6$. Consider the - sign in Eq. (1) that  gives a  barrier ($(V(0)=135.5+0i)$
for the Hermitian case, an absorptive-barrier for  $\nu=-0.6-2i$ $(V(0)=135.14-14.4i$ (see Fig. 2(a)) and an emissive barrier for
$\nu=-0.6+2i$ $(V(0)=135.14+14.4i)$ (see Fig. 2(b)). When we consider the + sign in Eq. (1) that gives a real well $(V(0)=-135.14+0i)$
for the Hermitian case, an emissive well for  $\nu=-0.6-2i$ $(V(0)=-170.14+14.4i)$ (see Fig. 2(c)) and an absorptive well
for $\nu=-0.6+2i$ ($V(0)=-170.14-14.4 i$) (see Fig. 2(d)). When the potential is non-absorptive $R(E)$ is more than that of the Hermitian case.
In Fig. 2(b) a spectral singularity co-exists with the deep oscillations in the reflectivity, see a  peak at $E=152.723$ in the dark curve. 
Here the non-Hermitian potential is again  non-absorptive. 

In Fig. (3), we present the scenario ($\{17\}$ in Table 1) of co-existence of spectral singularity and deep oscillations in the reflectivity.
In Fig. 3(a), the solid curve gives the reflectivity for the non-Hermitian case with a spectral singularity at $E=24.01$ followed by minima.
The real part of the potential is a well with side barriers (see the dark curve in Fig. 3(b)) and the imaginary part is again positive definite (light curve scaled up  by 
a factor of 10). The Figs. 3(c,d), like Figs 1(c,d) negate a possibility of a second SS in this case too.

For a complex non-Hermitian potential we  have the non-unitarity of ther flux resulting in $U(E)=R(E)+T(E)\ne 1$. Despite non-Hermiticity the discrete energies where $U(E)$ becomes 1 are of importance. However, presently only the cases
where $R(E)=0, T(E)=1$ have been termed the points of
invisibility [15,16]. In this regard we find that the deep minima in $R(E)$ (see Figs. 2,3) may  correspond  at the best to $T(E)\sim .96$ or so but not to 1. $T(E)$
becomes 1 only when the imaginary part is switched
of and the Ginocchio potential becomes  both reflectionless $(R(E)=0)$ and  invisible irrespective of the direction of incidence. In fact, such reflectionlessness
has been very well studied in Hermitian quantum mechanics
in the past [21]. Eventually, we find an essential absence of invisibility in complex Ginocchio potential. Nevertheless,
this potential being symmetric and entailing deep minima in $R(E)$ displays minimal visibility at selected energies without handedness.

Let us now study the behaviour of the scattering solution of the Ginocchio's potential (1). One of its solutions which does not diverge at 
$x=\infty$ is written as 
\begin{eqnarray}
&\Psi(x)=[\lambda^2+(1-\lambda^2)z^2]\left({1-z^2 \over 4}\right)^{-ik/(2\lambda^2)} ~_2F_1[\Omega(E), \Delta(E),1-ik/\lambda^2; {1-z \over 2}], 
\\ \nonumber
&z={\lambda y(x)\over \sqrt{1+(\lambda^2-1)y^2(x)}}.
\end{eqnarray}
The function $~_2F_1[p,q,r;w]=1+{pq \over r}{w \over 1!}+ {p(p+1)q(q+1) \over r(r+1)}{w^2\over 2!}+ {p(p+1)(p+2)q(q+1)(q+2) \over r(r+1)(r+2)}{w^3 \over 3!}+...$
is called Gauss hypergeometric function. When $p$ or $q$ or both become zero or negative integers,  $~_2F_1[-i,-j,r;w]$ becomes a polynomial of $w$. 
Notice that in all the cases in the Table 1, $\Delta(E_*)=n(\mbox{integer} \le 0)$, specially in the case $\{17\}$ (Table 1), $\Delta(E_*)=0$, we have $~_2F_1$ identically equal to 
1. Using (2) it can be shown  that when $x \sim \infty, z(x) \sim [1-{2e^{2\lambda^2(r_0-x)} \over \lambda^2}]$. In this situation $~_2F_1[p,q,r;1-z] \approx 1$ and 
$(1-z^2)/4 \sim (1-z)/2 \sim C(E) e^{ikx}$. The wavefunction (8) represents a transmitted wave traveling to right. Similarly using the properties of
$~_2F_1$, it turns out [19] that the wave function for $x \rightarrow -\infty$ would represent a linear combination of two (incident and reflected) plane waves :
$A(k) e^{ikx}+ B(k) e^{-ikx}$. Therefore, in all the cases of spectral singularities presented in the Table 1 the wavefunction (8) essentially represents scattering states on either side of
the origin.

Here $A(E),B(E), C(E)$ are Jost functions giving rise to  $R(E)=|{B(E) \over A(E)}|^2$ and $T(E)=|{C(E) \over A(E)}|^2$. At a spectral
singularity it is $A(E)$ that vanishes ($A(E_*)=0$) resulting in infinite  reflection and transmission. In Hermitian quantum mechanics this usually happens 
at a complex energy $(E_n-i\Gamma_n/2)$. These complex energies $E_n$ represent resonances (Gamow-Sigert states) and $\Gamma_n$ give the width of the 
resonance. But in the case of non-Hermitian potentials spectral singularity at a real energy $E_*$ signifies zero-width resonance.

To conclude, we have brought out new features in the versatile Ginocchio's potential (1) when it is made non-Hermitian.
These features are the existence of at most one spectral singularity and deep multiple minima in the reflectivity.
These features mostly require thick (wide) well/barrier  and hence $\lambda>1$.
It turns out that we get spectral singularity only when the potential is emissive (no-absorptive). For an absorptive potential
the spectral singularity can exist in the time reversed reflectivity and transmittivity namely, $R(-k)$ and $T(-k)$.
Besides this,  an absorptive potential entailing a spectral 
singularity is yet to be found.
We find that the reflectionlessness of this Hermitian potential gives way to deep oscillations 
in the reflectivity when the potential is perturbed to become non-Hermitian. More interestingly the first
reflectivity zero of the real Hermitian potential changes over to a spectral singularity of the (perturbed)
non-Hermitian potential. This feature requires more investigations to find out whether it is general
 and whether only the first reflectivity zero undergoes this transition. The reflectivity minima found
 here represent minimal visibility in any case whether the wave incident from left or right. 
Though Ginocchio potential has been made complex PT-symmetric in the past [22]. However there are
complications in studying the spectral singularity(s) in it. 
This  remains to be done next.

\section*{References}
\begin{enumerate}
\item C. M. Bender and S. Boettcher 1998, Phys. Rev. Lett. {\bf 80}, 5243.
\item Z. Ahmed 2001, Phys. Rev. A {\bf 64}, 42716.
\item Z. Ahmed 2004, Phys. Lett. A {\bf 324}, 152.
\item R. N. Deb, A. Khare, B.D. Roy 2003, Phys. Lett A {\bf 307}, 225. 
\item G. Levai, F. Cannata and A. Ventura 2001, J. Phys. A: Math. Gen. {\bf 34}, 839.   
\item F. Cannata, J.-P. Dedonder, and A. Ventura 2007, Ann. Phys.(N.Y.) {\bf 322}, 397.
\item B. Samsanov 2005, J. Phys. A: Math. Gen. {\bf 38}, 2571.    
\item A. V. Sokolov, A.A. Andrianov, F. Cannata 2006, J.Phys. A: Math. Gen. 10207. 
\item A. Mostafazadeh 2009, Phys. Rev. Lett. {\bf 102}, 220402. 
\item A. Mostafazadeh and H. Mehr-Dehnavi 2009, J. Phys. A {\bf 42}, 125303 (2009);
 A. Mostafazadeh 2009, J. Phys. A: Math. Theor. {\bf 44}, 375302.
\item S. Longhi 2009, Phys. Rev. Lett. {\bf 103}, 123609.
\item Z. Ahmed 2009, J. Phys. A: Math. Theor. {\bf 42}, 473005. 
\item B. Bagchi, C. Quesne 2010, J. Phys. A: Math. Theor. {\bf 43}, 305301.
\item Z. Ahmed 2012, J. Phys. A: Math. Theor. {\bf 48} 032004.
\item Zin Lin, Hamidreza Ramezani, Toni Eichelkraut, Tsampikos Kottos, Hui Cao, and Demetrios N. Christodoulides
2011, Phys. Rev. Lett. {\bf 106}, 213901.
\item A. Mostafazadeh 2012, math-ph 1206.0116v1. 
\item A. Mostafazadeh 2012, quant-ph 1205.4560v1.
\item A. Sinha 2012, Eur. Phys. Lett. vol. {\bf 98}, 60005. 
\item J. N. Ginocchio 1984, Ann. Phys. {\bf 152}, 203.
\item B. Sahu, Sk. Agarwalla, C. S. Shastry 2002, 
J. Phys. A: Math. Gen. {\bf 35}, 4349; B. Sahu and B. Sahu
 2009, Phys. Lett. A {\bf 373}, 4033.
\item M.V. Berry 1982, J. Phys. A: Math. Gen. {\bf 15}, 3691; 
J.D. Chalk 1988, Am. J. Phys. {\bf 56}, 29;
L.V. Chebotarev 1995, Phys. Rev. A{\bf 52}, 107;
Z. Ahmed 1996, Phys. Lett. A {\bf 210}, 1;
Z.Ahmed,C.M. Bender, M.V.Berry 2005, J. Phys. A: Math. Gen. L627.;
Z.Ahmed 2006, J. Phys. A: Math. Gen. {\bf 39}, 7341.
\item G. Levai, A. Sinha and P. Roy 2003, J. Phys. A: Math. Gen. {\bf 36}, 7611. 
\end{enumerate} 
\begin{table}[h]

	\centering
		\begin{tabular}{|c||c||c|c|c|c|c|c|}
		\hline
		S.N  & Sign& $\nu$& $\lambda$& $E_*$& $V(0)=\mp \lambda^2\nu(\nu+1)$& $\Delta(E_*)$&$Re(V(x))$\\
		&  Eq.(1) & & & & + $(1-\lambda^2)/2$& $=n$ & \\
		\hline
		\hline
		$\{1\}$&-&$-2.65i$	& 3.4 & 104.229 & $70.90+30.63i$ & -1& repulsive barrier\\
		\hline	
		$\{2\}$&-&$1-4.5i$ & 4.123 & 650.126 & $302.235 + 229.488 i$ & -4& repulsive barrier\\
		\hline 
		$\{3\}$&-& $6-12i$ & 1.4366 & 359.557& $209.978+322.359i$& -8& repulsive barrier\\
		\hline
		$\{4\}$&-& $-8.39+10.4i$ & 1.351 & 248.522 & $83.8348+299.537i$ & -9 & repulsive barrier \\
		\hline
		$\{5\}$&-& $4.2-12.57i$ & 1.2 & 239.275 & $195.857+170.148i$ & -5 & repulsive barrier\\
	  \hline
	  $\{6\}$&-& $-6.99+11.3i$& 1.0 & 127.690 & $85.890+146.674i $ & -6 & repulsive barrier \\
	  \hline
	  $\{7\}$&+&$ 3.75+.5i$ & 3.1221 & 190.868 & $166.817+41.4269i$ & -1 & repulsive barrier\\
		\hline
		$\{8\}$ & +& $2.776+2.15i$ & 2.1 & 78.761 & $24.1326+62.122i$ & -3 & repulsive barrier  \\
		\hline
		$\{9\}$&+ & $-7.384-3.05i$ & 1.63 & 153.668 & $ 99.706+111.570i$ & -4 & repulsive barrier\\
		\hline
		$\{10\}$&+& $-4.75-4.928i$ & 1.85 & 121.598 & $-23.364+143.362i$ & -6 & attractive well \\
		\hline
		$\{11\}$ & + & $4.67+ 7.8366i$ & 1.74 & 166.720 & $-106.778+ 245.328i$ & -9 & attractive well \\
		\hline
		$\{12\}$&-& $ -6+i$ & 4.261 & 236.028 & $-535.106+199.717i$ & -6& attractive well \\
		\hline
		$\{13\}$&+&$-3-8.5i$ & .96 & 5.3077 & $-61.016 + 39.168 i$ & -8 & attractive well\\
		\hline
		$\{14\}$&-& $.55-.6i$ &  9.312 & 478.444 & $-85.563 + 109.44i $ & -2& well with side barriers\\
		\hline
		$\{15\}$&+& $-1.560-.601i$ & 10 & $651.183$ & $1.6769+127.572 i$ & -2 & well with side barriers\\
		\hline
		$\{16\}$&-& $1.9-2.4i$ & 2.127 & 55.4231 & $-0.6310+52.118i$ & -3& well with side barriers\\
		\hline
		$\{17\}$&-& $-.6+.5i$ & 7 & 24.01 & $0.01+4.9 i$ & 0 & well with side barriers\\
		\hline
		$\{18\}$&+& $-.6-3.4i$ & 4.5 & 3.9130 & $-248.575+13.771 i$ & -3 & well with side barriers \\ 
		\hline
		$\{19\}$ &+& $a-ib,a,b>0$ & & no SS & $V_1-iV_2, V_2 >0$ & & \\
		\hline
		$\{20\}$ &-& $a+ib,a,b>0$ &  &  no SS & $V_1-iV_2, V_2>0$ & &  \\
		\hline
		
		\end{tabular}
		\caption
		{Occurrence of the spectral singularity in Ginocchio's complex non-Hermitian potential [Eq.(1)]. 
Various details of the cases $\{11\}, \{17 \}$ are depicted 	in Fig. 1 and 3, respectively.	
In all the cases, the SS occurs when the imaginary part is positive (emissive) and does not seem to occur when it is negative (absorptive).  Here the cases (see $\{19,20\}$) underline this observation.} 
		\end{table}

\begin{figure}
\centering
\includegraphics[width=16 cm,height=14 cm]{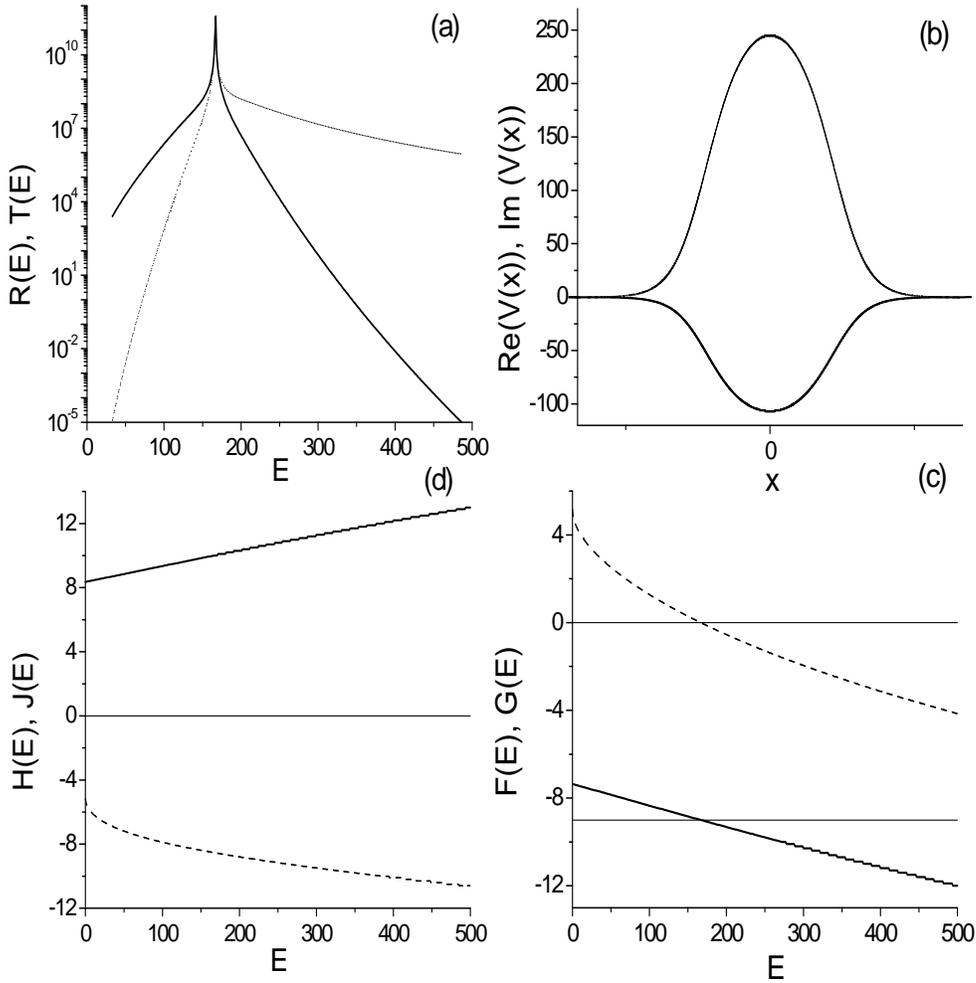}
\caption{A typical scenario of the spectral singularity(SS) in the complex Ginocchio's potential [Eq.(1)]. See the parameters for $\{11\}$ in Table 1 where $\nu=4.67+7.8366i$. (a): $R(E)$ (dark curve) and $T(E)$ (light dotted curve) both become as large as $(~10^{11})$ at the SS $E=E_*=166.72$.
(b): The corresponding potential profile $V(x)$ [Eq.(1)]: the real part (dark curve) and imaginary part (light curve). 
(c): See Eq. (4), the function $G(E)$(dashed curve) is monotonically 
decreasing and cutting the energy axis at $E=E_*$.  The function $F(E)$ (solid curve) is monotonically decreasing and cutting  the line $y=-9$ at $E=E_*$. These two simultaneously fulfill the condition (6) $\Delta(E)=n$ giving rise to very large values of  $R(E_*)$ and $T(E_*)$ and hence the spectral singularity. On the contrary in (d), notice that $J(E)$ (dotted curve) [see Eq.(7)] remains decreasing and negative whereas $H(E)$ (solid curve) is increasing and positive ruling out the existence of a second SS.}  
\end{figure}

\begin{figure}
\centering
\includegraphics[width=16 cm,height=14 cm]{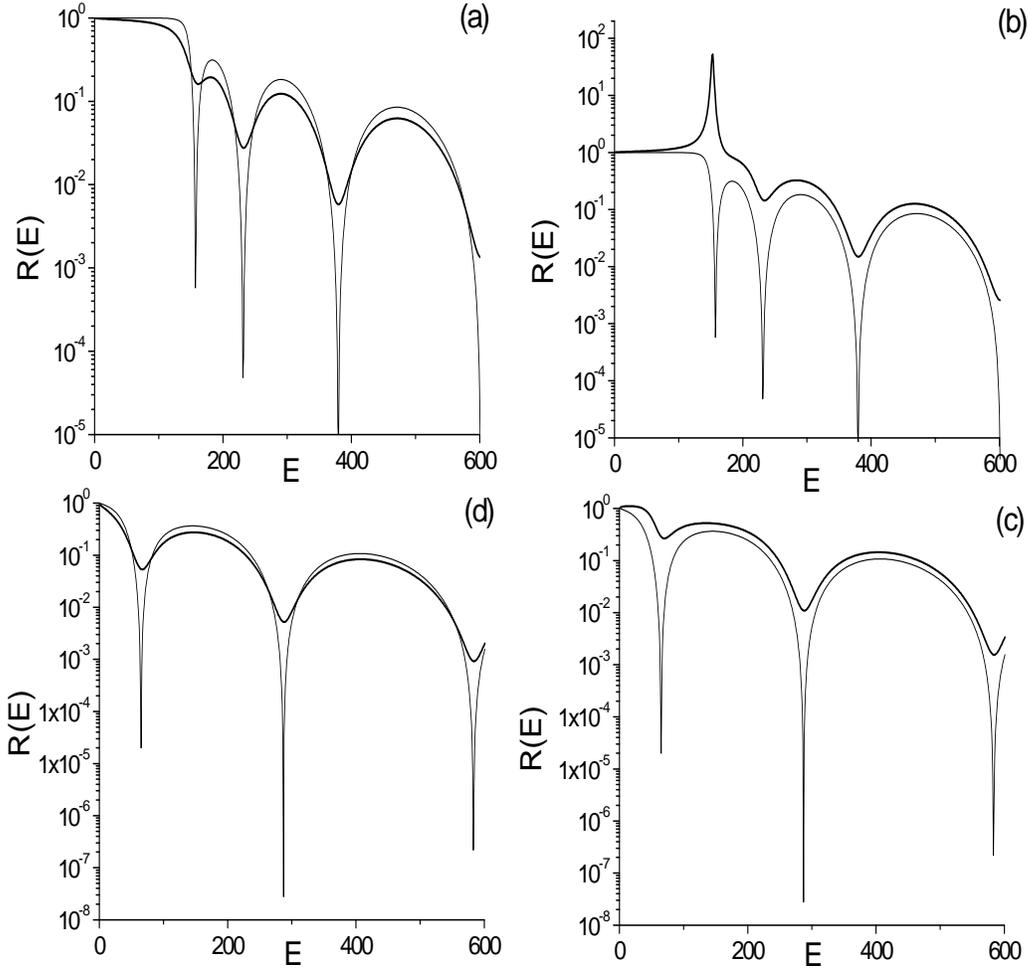}
\caption{A typical scenario of the reflectionlessness of the Hermitian case ($\nu=-1/2\pm 2i$, light curves) 
and the energy oscillations in the reflectivity for the non-Hermitian case (dark curves) when $Re(\nu) \approx -0.5$
and $Im(\nu)$ is small. Smaller the $Im(\nu)$  sharper are the reflectivity-minima, but here we take a moderate value of 2 and $\lambda=6$. The parts (a-d) correspond to
various cases of $\nu=-.6 \pm ib$  and $\pm$ signs in Eq. (1)
(see the text). The part (b) displays that a spectral singularity co-exists with the deep oscillations in the reflectivity, see a  peak at $E=152.723$ in the dark curve. 
Here the non-Hermitian potential is again  non-absorptive.}
\end{figure}

\begin{figure}
\centering
\includegraphics[width=16 cm,height=14 cm]{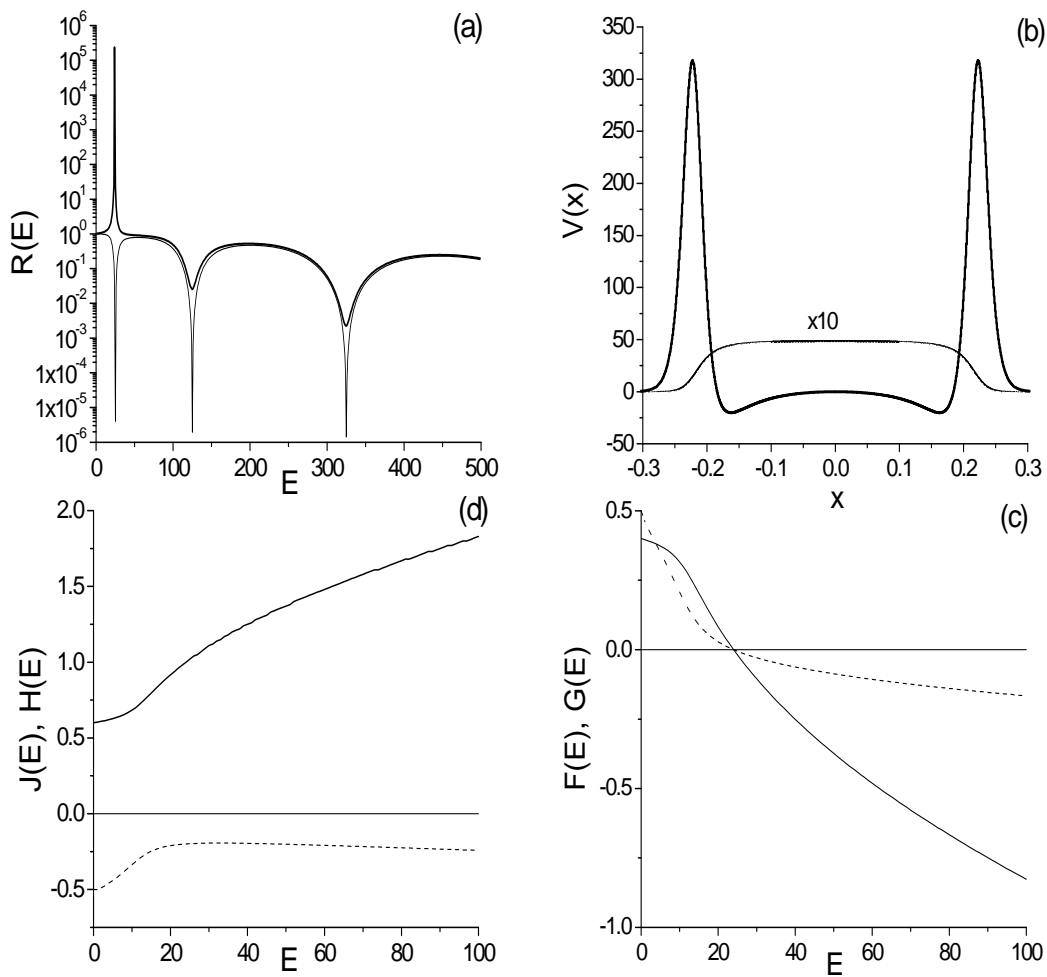}
\caption{(a): The scenario of the co-existence of a spectral singularity and deep minima in the reflectivity for the case of $\{17\}$ in Table 1.
The light curve is for the Hermitian case $(\nu=-.5+.5i, V(0)=.5+0i)$ and the dark curve is for the non-Hermitian case $\nu=-.6+.5i, $. 
The value of $\lambda=7$ is common for both. (b):  The real part
of the potential is a well with side barrier shown by dark curve, the light curve represents the imaginary part of $V(x)$ scaled up by a factor of 10.
(c,d) same as Fig. 1. Notice that at $E=24.01$ we get $F(E)=0=G(E)$, so an SS as $\Gamma(0)=\infty$.}

\end{figure}

\end{document}